\documentclass[journal]{IEEEtran}

\usepackage{cite}
\usepackage[colorlinks=true,
            linkcolor=blue,
            citecolor=blue,
            urlcolor=blue]{hyperref}
\usepackage{algorithmic}
\usepackage{graphicx}
\usepackage{textcomp}
\usepackage{xcolor}
\usepackage{booktabs}
\usepackage{tikz}
\usetikzlibrary{arrows.meta,positioning}
\usepackage{listings}
\usepackage{url}

\title{A Measurement Plane for Quantum Networking}

\author{
Abderrahim Amlou$^{1,2}$,
Amar Abane$^{1}$,
Anouar Rahmouni$^{1}$,
Mheni Merzouki$^{1}$,
Abdella Battou$^{1}$,
Ahmed Lbath$^{2}$,
Ya-Shian Li-Baboud$^{1}$,
Oliver Slattery$^{1}$,
and Thomas Gerrits$^{1}$
\thanks{\textit{$^{1}$National Institute of Standards and Technology, Gaithersburg, MD.}}%
\thanks{\textit{$^{2}$Univ. Grenoble Alpes, Grenoble.}} 
\thanks{\textit{Corresponding author: Abderrahim Amlou, abderrahim.amlou@nist.gov.}}
\thanks{This work has been submitted to the IEEE for possible publication.
Copyright may be transferred without notice, after which this version may
no longer be accessible.}
}

\begin{document}
\maketitle

\begin{abstract}
Quantum networking testbeds lack a distinct plane for coordinating distributed measurements and collecting experimental data across heterogeneous devices. To address this gap, we present the \emph{Measurement Plane}, a dedicated plane that complements the data, control, and management planes rather than replacing or extending their pipelines. The contribution is presented as a distributed framework that organizes measurement functions into four layers: application, experiment coordination, capability, and resource agents. Our design separates user workflows from device-specific control.

We implemented the framework as containerized microservices connected through publish--subscribe messaging, and validated it on a two-node quantum networking setup connected by an optical network. The framework successfully coordinated remote nodes to execute coincidence measurement and polarization entanglement distribution experiments with visibility interference of up to 98 percent. This evaluation demonstrated the effectiveness of the framework for supporting complex, distributed quantum experiments, enabling online measurement and feedback, and significantly reducing manual configuration and execution effort.

\end{abstract}


\section{Introduction}

Quantum networking testbeds are evolving from isolated experimental demonstrations into integrated infrastructures capable of supporting distributed quantum applications \cite{islam2025experimental}. These systems require not only a quantum data plane for entanglement distribution and quantum state manipulation \cite{abane2025entanglement, kozlowski2023rfc}, but also a control and measurement infrastructure that enables researchers to operate, calibrate, and characterize quantum hardware in a reproducible and scalable manner \cite{bersin2024development}. As quantum network testbeds expand to support multiple users, diverse hardware components, and increasingly complex experiments, the need for a structured Measurement Plane becomes critical.

Current approaches for quantum testbed operation typically embed measurement capabilities within device-specific control software and often rely on loosely structured architectures or ad hoc scripting to coordinate experiments across multiple hardware components~\cite{monga2025quantnet}. While sufficient for small-scale or single-experiment demonstrations, these approaches face significant limitations as testbeds grow in complexity. These approaches tightly couple experimental workflows to hardware implementations, complicate reuse and shared access, and make distributed, multi-step procedures difficult to automate. Moreover, conventional control frameworks are not designed for continuous and high volume experimental data delivery.

This work addresses these limitations by treating measurements as an independent plane in quantum network testbeds. Rather than embedding measurement procedures within the control plane or distributing them across device-specific tools, our design organizes measurement capabilities into a set of architectural layers with unified interfaces and coordination. The Measurement Plane operates alongside the data, control, and management planes \cite{abane2025entanglement, kozlowski2023rfc}, exposing hardware capabilities through a capability-based abstraction that decouples measurement requests from device-specific implementation details.

This capability-based abstraction enables measurement functions to be discovered, invoked, and composed in a uniform manner across the testbed. In addition, the Measurement Plane also enables the coordination of complex experimental workflows. As a result, experimental procedures can be automated, reused, and executed consistently across different experimental deployments.


We deploy and validate the proposed Measurement Plane on a quantum networking testbed that connects two nodes separated by approximately 1.4~km of optical fiber. Through this deployment, we demonstrate the ability of the Measurement Plane to coordinate distributed measurement resources and automate complex experimental procedures. In particular, we show how the system enables polarization entanglement distribution by performing coincidence detection measurement across remote nodes, and supports automated polarization fringe measurements through capability composition and experiment workflow execution.

\section{Related Work}
\label{sec:related}

The proposed Measurement Plane lies at the intersection of two research directions. Quantum-network architectures provide mechanisms for managing and controlling distributed quantum hardware, scheduling operations, and coordinating protocols, but measurement functions are generally embedded within control, service, or device-specific software. Classical measurement platforms instead treat measurement as a programmable and reusable service, but they are not designed for the hardware coordination and dependent workflows required by quantum experiments. This section reviews these two directions and identifies the architectural gap addressed by our work.

\subsection{Quantum Network Operation and Control}

Existing quantum-network architectures primarily focus on operating the network and coordinating quantum resources. The QUANT-NET project introduced a two-level control framework that separates network-wide coordination from node-level, low-latency control \cite{monga2025quantnet}. In fact, its QUANT-NET Control Plane (QNCP) provides resource-management primitives, scheduling mechanisms, and extensible interfaces for coordinating time-slotted quantum operations and calibration procedures \cite{yu2025extensible,monga2025quantnet}.

The ArQNet coordinator similarly applies software-defined networking principles to organize the Argonne Quantum Network into infrastructure, control, and service planes \cite{islam2025experimental}. Its control infrastructure supports resource allocation, scheduling, topology management, and continuous entanglement generation across a distributed testbed. At the node level, QNodeOS provides an operating-system abstraction for executing and scheduling quantum-network applications on heterogeneous quantum processors \cite{delle2025operating}.

These systems demonstrate that measurement, calibration, and hardware characterization are essential to quantum-network operation. However, such functions are generally implemented as parts of control logic, service execution, or node-specific software. They are not exposed through a unified, network-wide abstraction through which measurements can be discovered, invoked, composed, and shared independently of the underlying devices.

\subsection{Distributed Measurement Infrastructures}

Classical networking research provides a complementary perspective in which measurement is treated as a distributed service. Platforms such as perfSONAR and RIPE Atlas deploy measurement functions across distributed infrastructure and provide common interfaces for initiating measurements and collecting results \cite{perfsonar2024,ripeatlas2024}. These systems demonstrate the value of separating measurement requests from the probes and services that implement them.

The mPlane architecture extends this idea through a capability-based model in which measurement components advertise available functions and clients invoke them through structured specifications \cite{mplane2014}. CaSpeR adopts a similar capability model and uses publish--subscribe messaging to support loosely coupled measurement services and scalable result distribution \cite{abane2023data}. These approaches provide useful mechanisms for capability discovery, common request formats, asynchronous communication, and shared access to measurement results.

However, classical measurement platforms typically collect independent observations such as connectivity, traffic, or performance metrics. Quantum-network experiments impose additional requirements: measurements must often be synchronized with state preparation and hardware configuration, multiple distributed devices may participate in a single measurement, and the output of one phase may determine the parameters or execution of subsequent phases. Consequently, quantum measurements require both distributed data collection and experiment-level coordination.

\subsection{Position of This Work}

The reviewed approaches address complementary parts of the problem. Quantum-network architectures provide domain-specific control, scheduling, and hardware coordination, whereas classical measurement platforms provide reusable measurement abstractions and distributed data-delivery mechanisms. The remaining gap is an architecture that combines these properties for distributed quantum-network experiments.

The Measurement Plane addresses this gap by treating measurement as a first-class architectural function that operates alongside the data, control, and management planes. It exposes measurements through reusable capabilities, separates device access through resource agents, and coordinates capabilities through multi-step experimental workflows. In this way, the framework retains the distributed and programmable characteristics of classical measurement infrastructures while supporting the hardware interaction, timing dependencies, and feedback required by quantum-network testbeds.

\section{Measurement Plane Architecture}
\label{sec:architecture}
This section presents the Measurement Plane architecture for quantum network testbeds. We first summarize the design goals that motivate the architecture, then describe its deployment model and four functional layers. The layer description is presented from the experimental hardware upward, using coincidence measurement and polarization-fringe analysis as running examples. We then describe the capability interface, measurement lifecycle, and identifier hierarchy used to invoke and track measurements.


\subsection{Overview and Design Rationale}
As mentioned above, measurement is treated as a distinct plane of the quantum network architecture that operates alongside the data, control, and management planes. The user remains outside this plane and interacts with it through the application layer, while the underlying testbed also remains outside the plane and is reached only through the resource agent layer. Within the Measurement Plane, communication follows a layered structure in which each layer interacts only with the adjacent ones.

The architecture is designed around several practical goals motivated by how users interact with the quantum network testbed:
\begin{itemize}
    \item First, the system is modular. New capabilities should be introduced without changing how users interact with the Measurement Plane. From the user perspective, submitting an experimental request should always follow the same pattern: select a capability or workflow, choose the target endpoint when needed, and provide the required parameters through the application interface.

    \item Second, capabilities are reusable and composable. A capability should not only provide a specific function but also serve as a building block for constructing more advanced capabilities. Users should be able to implement higher-level functionalities by coordinating multiple existing capabilities or extending them (e.g., with additional processing or metrics). This composability simplifies the development of new measurement features and avoids duplication of low-level implementations.

    \item Third, the architecture supports concurrent access. Multiple users or services can invoke the same capability simultaneously on different endpoints. They can also subscribe to the results of the same active measurement without requiring separate executions or manual coordination.

    \item Finally, the architecture supports the automation of complex procedures. Rather than relying on large collections of device-specific scripts, users should be able to describe multi-step experimental procedures at a higher level. These procedures can then be orchestrated and executed using appropriate capabilities.
\end{itemize}

Another important design property is support for distributed deployment. Resource agents run close to the hardware they manage, so device interaction and raw-data acquisition remain local to the corresponding experimental nodes. Capabilities and coordination services may be deployed locally or centrally, depending on their processing and orchestration requirements. Components communicate through a shared messaging infrastructure, allowing devices, capabilities, and services to be distributed across multiple locations without requiring direct point-to-point integration.

\begin{figure}[!t]
    \centering
    \includegraphics[width=0.72\columnwidth]{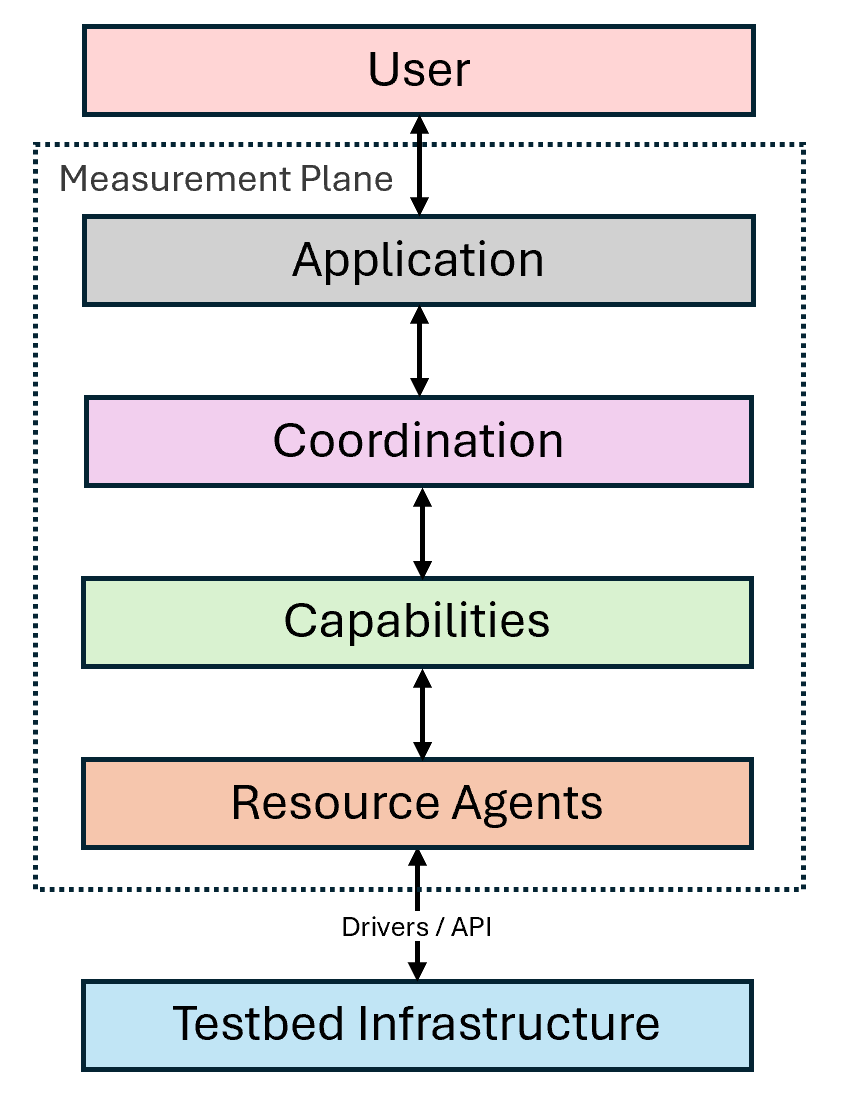}
    \caption{Layered architecture of the Measurement Plane.}
    \label{fig:mp_architecture_overview}
\end{figure}

\begin{figure}[!t]
    \centering
    \includegraphics[width=\columnwidth]{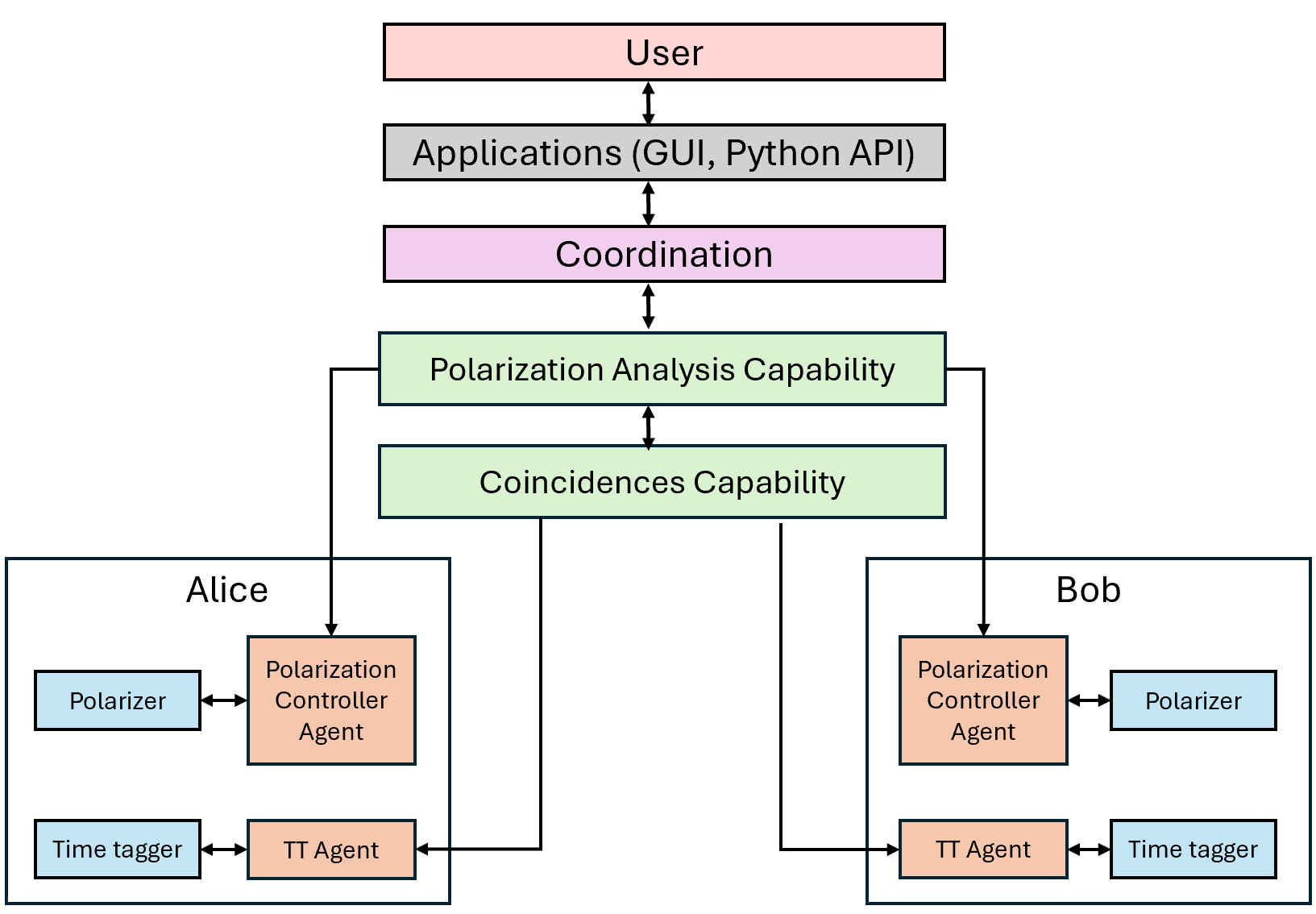}
    \caption{Software interaction model of the Measurement Plane during experiment execution.}
    \label{fig:software_interraction}
\end{figure}

The Measurement Plane is organized into four functional layers, as illustrated in Fig.~\ref{fig:mp_architecture_overview}. To make the role of each layer concrete, we describe the architecture from the testbed infrastructure upward using two examples. The first is a coincidence measurement that combines photon-detection timestamps collected at two remote nodes to assess the correlation between detection events. The second is a polarization fringe experiment in which analyzer settings are varied and the coincidence measurement is repeated at each setting. Figure ~\ref{fig:software_interraction} presents the software interaction model of the Measurement Plane for these two examples. The figure shows how the time-tagger (TT) and polarization-controller are exposed from the Alice and Bob nodes to the user through the Measurement Plane.

\subsection{Resource Agent Layer}

The resource agent layer forms the boundary between the Measurement Plane and the experimental testbed. A resource agent runs close to the hardware it manages and communicates with the device through its driver or application programming interface (API). It translates device-specific operations and data formats into a common messaging interface that can be used by the upper layers.

This function can be illustrated using the two running examples introduced earlier. In the coincidence-measurement example, one \textit{time-tagger (TT) resource agent} runs at each remote node. Each agent acquires photon-arrival timestamps from its local time tagger and makes them available to the Measurement Plane. Although this TT device may rely on different vendor-specific software, it appears to the rest of the Measurement Plane through the same abstraction mechanism.

As a result, higher layers interact with resources through a uniform interface rather than through device-specific commands. This separation not only enables the use of heterogeneous hardware within the same experimental workflow, but also simplifies extensibility. Supporting a new vendor implementation for an existing resource type typically requires only the addition of the corresponding driver, while the resource agent and the layers above it can remain unchanged.

\subsection{Capability Layer}

The capability layer transforms device access into reusable measurement and analysis functions. A capability represents a well-defined operation together with its accepted parameters, produced results, and relevant metadata. It may use one or more resource agents while remaining independent of their device-specific interfaces. A capability may also build upon other capabilities, enabling functionality to be composed hierarchically.

As shown in Fig.~\ref{fig:software_interraction}, the \textit{coincidence capability} obtains timestamp streams from the \textit{time-tagger resource agents} at the two remote nodes. It aligns the streams and computes coincidence counts or a coincidence histogram. From the perspective of this capability, the time taggers are accessed through the same resource-agent interface regardless of their vendor or local driver.

In addition, the same \textit{coincidence capability} can be reused by a higher-level \textit{polarization analysis capability}. This capability takes polarization angles as input, configures the corresponding polarizers through the \textit{polarization controller resource agents}, and then invokes the \textit{coincidence capability} to obtain the measurement results between Alice and Bob. By combining lower-level capabilities in this manner, more advanced measurement and analysis functions can be implemented without duplicating existing logic.

\subsection{Experiment Coordination Layer}

Next, the experiment coordination layer organizes capabilities into complete measurement procedures. It receives an experimental request and determines which capabilities must be invoked, in what order, and with which parameters. The coordinator manages the execution state and transfers results between dependent steps, but it does not communicate with hardware directly.

For example, a coincidence measurement is a simple workflow. The coordinator invokes the \textit{coincidence analysis capability} with the selected nodes, acquisition duration, and analysis parameters, and then returns the resulting histogram or coincidence count to the upper layer.

In contrast, the polarization fringe experiment requires a more complex workflow. The coordinator controls the sweep over the experiment settings. For each setting, it invokes the \textit{polarization analysis capability}, which uses the \textit{polarization controller resource agents} and \textit{coincidence analysis capability} to produce one measurement point. The coordinator repeats this process over the requested range of angles, aggregates the results produced at each setting, and returns the complete fringe dataset to the upper layer.

Finally, experiments may be described declaratively, for example through a YAML specification that defines their phases, steps, parameters, and dependencies. The coordinator interprets this description and ensures that each step is executed in the correct order. This allows multi-phase and multi-step experiments to be automated without embedding the experimental logic directly into the application layer.

\subsection{Application Layer}
The application layer is the user-facing entry point to the Measurement Plane. It exposes the system through a graphical interface for interactive use and through a programmatic client library for scripts, notebooks, and external services. The user remains outside the Measurement Plane and interacts with it only through this layer.

This layer hides internal details such as message topics, serialization formats, capability definitions, and device-specific interfaces. Instead, it allows users to describe what they want to run, either as a single measurement invocation or as a multi-step experiment description.

For a coincidence measurement, the user selects the two remote endpoints and provides parameters such as the acquisition duration and coincidence window. For more complex experiments, the user defines an experiment description that specifies the sequence of measurements and their associated parameters, such as an analyzer-angle sweep for a polarization-fringe experiment. The application layer packages these inputs into structured requests, submits them to the coordination layer, and presents status updates and results back to the user.

\subsection{Capability Interface and Measurement Lifecycle}
\label{sec:capability}
This subsection describes how capabilities are represented, invoked, and tracked during execution. Communication within the Measurement Plane is organized around a set of message types that describe capability discovery, invocation, execution progress, and result delivery. The central concept is the \emph{capability}, which defines a unit of functionality available to the coordination service through the Measurement Plane.

Each capability is advertised through a structured description that contains the information needed for discovery and invocation. This includes a human-readable label, the endpoint providing the capability, the capability name, the schema of accepted parameters, the schema of produced results, and additional metadata such as units, timing information, and execution semantics.

When the coordination layer invokes a capability, it sends a specification message containing the requested parameters. If the request is valid and can be admitted for execution, the capability returns a receipt. During execution, the capability may emit status messages to report progress and result messages to publish measurement outputs. Separating lifecycle information, such as receipts, progress updates, interruptions, and completion events, from measurement data improves observability while allowing clients to process result streams independently.

Capabilities may follow different execution modes. Some measurements are one-shot operations that return a single result, while others produce a finite stream or a continuous stream that remains active until interrupted. Internally, a capability may obtain data or perform control operations by interacting with lower-level capabilities or resource agents responsible for the underlying devices.

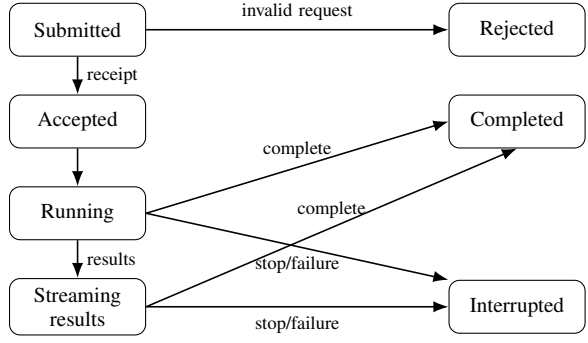
\begin{figure}[!t]
    \centering
    \begin{tikzpicture}[
        node distance=5mm and 40mm,
        state/.style={draw, rounded corners, align=center, minimum width=1.8cm, minimum height=7mm, font=\footnotesize},
        arrow/.style={-{Latex[length=2mm]}, semithick}
    ]
        \node[state] (submitted) {Submitted};
        \node[state, below=of submitted] (accepted) {Accepted};
        \node[state, below=of accepted] (running) {Running};
        \node[state, below=of running] (streaming) {Streaming\\results};

        \node[state, right=of submitted] (rejected) {Rejected};
        \node[state, right=of accepted] (completed) {Completed};
        \node[state, right=of streaming] (interrupted) {Interrupted};

        \draw[arrow] (submitted) -- node[right, font=\scriptsize] {receipt} (accepted);
        \draw[arrow] (submitted.east) -- node[above, font=\scriptsize] {invalid request} (rejected.west);
        \draw[arrow] (accepted) -- (running);
        \draw[arrow] (running) -- node[right, font=\scriptsize] {results} (streaming);

        \draw[arrow] (running.east) -- node[above, font=\scriptsize] {complete} (completed.west);
        \draw[arrow] (streaming.east) -- node[above, font=\scriptsize] {complete} (completed.south);
        \draw[arrow] (running.east) -- node[below, font=\scriptsize] {stop/failure} (interrupted.north west);
        \draw[arrow] (streaming.east) -- node[below, font=\scriptsize] {stop/failure} (interrupted.west);
    \end{tikzpicture}
    \caption{Measurement lifecycle within the Measurement Plane. After submission, a request is either rejected or admitted for execution. Running measurements may produce status updates and result streams until completion or interruption.}
    \label{fig:measurement_lifecycle}
\end{figure}

Figure~\ref{fig:measurement_lifecycle} summarizes the measurement lifecycle, from request submission and admission to execution, result streaming, and termination.

\begin{figure}[!t]
    \centering
    \includegraphics[width=\columnwidth]{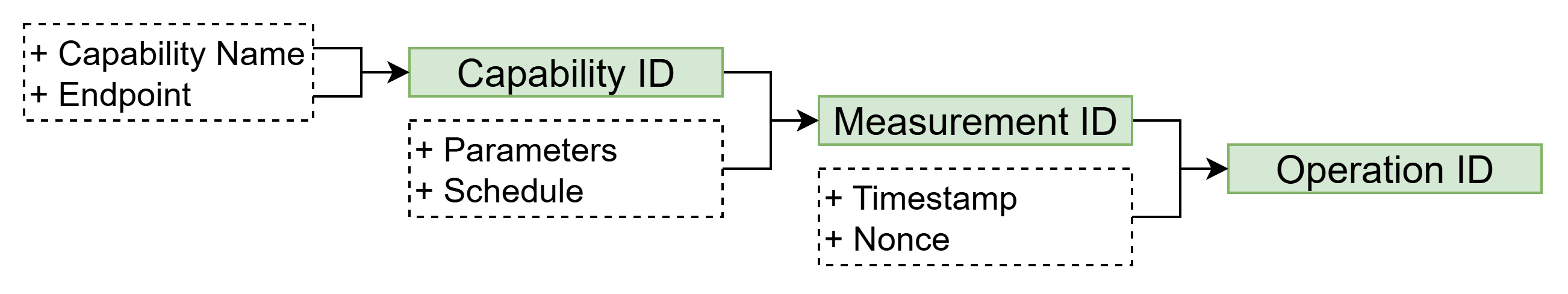}
    \caption{Identifier hierarchy used in the Measurement Plane.}
    \label{fig:id_hierarchy}
\end{figure}

This lifecycle is coupled with an identifier hierarchy that allows the system to route requests, share active measurements, and track client-specific invocations. The relationship between these identifiers is illustrated in Fig.~\ref{fig:id_hierarchy}.

The \textbf{\emph{Capability ID}} uniquely identifies a capability instance provided by a specific endpoint. While multiple endpoints may support the same capability type, each instance is associated with the endpoint that exposes it. The Capability ID therefore binds together the capability name and the endpoint providing it, allowing the system to route requests to the correct location.

The \textbf{\emph{Measurement ID}} identifies a concrete measurement definition. It is derived from the Capability ID together with the request parameters. Because this identifier is deterministic, two requests describing the same measurement produce the same Measurement ID. This allows different clients to independently compute the same identifier and subscribe to the same result stream.

If a capability receives a request whose Measurement ID matches an already active measurement, it does not need to start a second execution. Instead, it can attach the new requester to the existing execution and publish results to the same topic. This reduces unnecessary load on shared resources and allows multiple clients to observe the same measurement concurrently.

The \textbf{\emph{Operation ID}} provides a finer level of tracking by distinguishing individual invocations associated with the same measurement. It enables the system to track client-specific interactions with a shared execution without affecting the underlying measurement itself. If multiple clients request the same measurement, each invocation is associated with its own Operation ID. When one client stops its operation, the system acknowledges that request without interrupting the shared measurement as long as other clients remain subscribed to the results.

\section{Implementation}
\label{sec:implementation}
The Measurement Plane is implemented as a distributed microservice framework for experimental testbeds.

\subsection{System Overview}

The core logic of the Measurement Plane is implemented as a Python library that defines the messaging model, capability interfaces, and interaction patterns between system components. This library provides the abstractions required for capability advertisement, request handling, execution tracking, and result publication.

Individual system components, such as capabilities, resource agents, and the coordination service, are implemented as independent services using this library. In practice, each component runs as a containerized microservice. When the system is deployed, multiple services are started simultaneously using Docker, allowing the Measurement Plane to operate as a distributed system across several machines in the experimental network.

\subsection{Messaging Infrastructure}

Communication between components relies on a publish--subscribe messaging architecture implemented using the NATS messaging system. All components connect to the messaging broker and exchange messages through named topics corresponding to capability advertisements, specifications, receipts, status updates, and results.

Messages are encoded using JSON and follow schemas defined by the Measurement Plane library. This messaging infrastructure provides the transport used by the distributed services while keeping the implementation loosely coupled.

\subsection{Capabilities and Resource Agents Implementation}

Capabilities are implemented as Python classes that expose a specific measurement functionality. When deployed, each capability runs as an independent microservice that subscribes to specification messages corresponding to the capability it provides. Upon receiving a request, the capability validates the parameters using predefined JSON schemas and then executes the requested operation.

Resource agents are also implemented as containerized microservices deployed close to the corresponding hardware. Each agent exposes a uniform messaging interface while internally interacting with devices through Python drivers or the appropriate API. For example, the current implementation includes resource agents for time taggers and polarization controllers. The time tagger agent streams time-tag data obtained through a device driver, while the polarization controller agent allows setting or reading the angles of motorized wave plates.

\subsection{Experiment Coordination and Client Interfaces}

Experiment workflows are executed by a dedicated coordinator service implemented as a containerized microservice. This service receives experiment requests from the application layer and drives the execution of the corresponding workflow. Users interact with the Measurement Plane through two main interfaces: a Python client library for scripts, notebooks, and external services, and a graphical user interface (GUI) implemented as a web application.

The GUI communicates with the Measurement Plane using both REST and messaging interfaces. REST endpoints are used to submit requests, while the messaging infrastructure is used for online status updates and result streaming.


\section{Evaluation}
\label{sec:evaluation}
We validated the Measurement Plane with a polarization entanglement distribution experiment on a distributed quantum network testbed at the National Institute of Standards and Technology (NIST). We focused on two representative cases: coincidence detection and polarization fringe analysis. The goal is to verify coordinated execution and workflow automation across sites.

\subsection{Experimental Setup}

\begin{figure}[!t]
    \centering
    \includegraphics[width=\columnwidth]{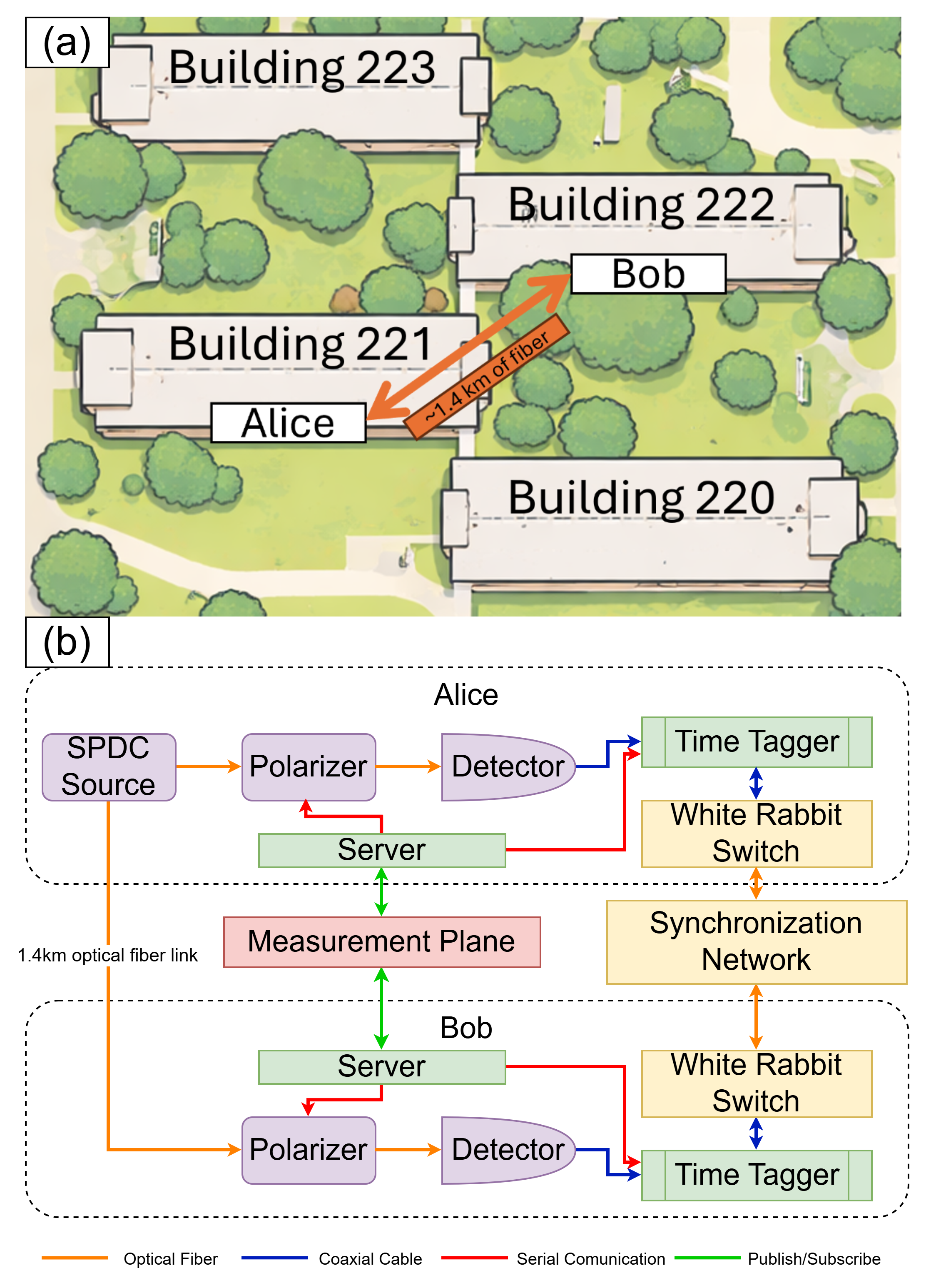}
    \caption{Experimental setup used to evaluate the Measurement Plane.
    (a) Physical deployment of the experiment across the NIST campus, showing the optical path between Building~221 (Alice) and Building~222 (Bob). The two nodes are connected through approximately $1.4\,\mathrm{km}$ of deployed optical fiber.
    (b) Experimental architecture at both nodes. At Alice, a spontaneous parametric down-conversion (SPDC) source generates photons that are measured locally and sent through the fiber link to Bob. Each node includes polarization analysis and single-photon detection, with time-tagged detection events collected by a time tagger synchronized through the White Rabbit network. Local servers host resource agents that interface the external testbed devices with the Measurement Plane and return measurement results to the higher layers.}
    \label{fig:experimental_setup}
\end{figure}

As illustrated in Fig.~\ref{fig:experimental_setup}, the evaluation was conducted across two nodes connected through the quantum testbed. The two nodes, referred to as Alice and Bob, are located in separate buildings connected by approximately 1.4~km of optical fiber.


An entangled photon source is located at the Alice node. One photon of each pair is detected locally at Alice, while the second photon is transmitted through the fiber link and detected at Bob. Each node is equipped with a single-photon detector connected to a time tagger responsible for recording photon arrival timestamps.

The nodes are synchronized using a White Rabbit timing system, providing a shared time reference for the distributed measurements. Each node hosts a local server running resource agents connected to the experimental hardware. In particular, two resource agents are deployed at each node: a time tagger agent responsible for acquiring timestamp data and a polarization controller agent responsible for adjusting the analyzer settings. 
The time-tag acquisition infrastructure implemented in the time tagger agents builds upon the scalable data acquisition system developed in~\cite{amlou2025scalable}.

Capabilities including coincidence analysis and polarization analysis, are deployed on a central server in the network together with the experiment coordinator and the graphical user interface. In the deployed system, experimental requests entered through the application layer, are orchestrated by the coordinator, executed by the relevant capabilities, and reached the hardware only through the resource agents.

\subsection{Coincidence and Polarization Measurements}

\begin{figure*}[!t]
    \centering
    \begin{minipage}[t]{0.48\textwidth}
        \centering
        \includegraphics[width=\textwidth]{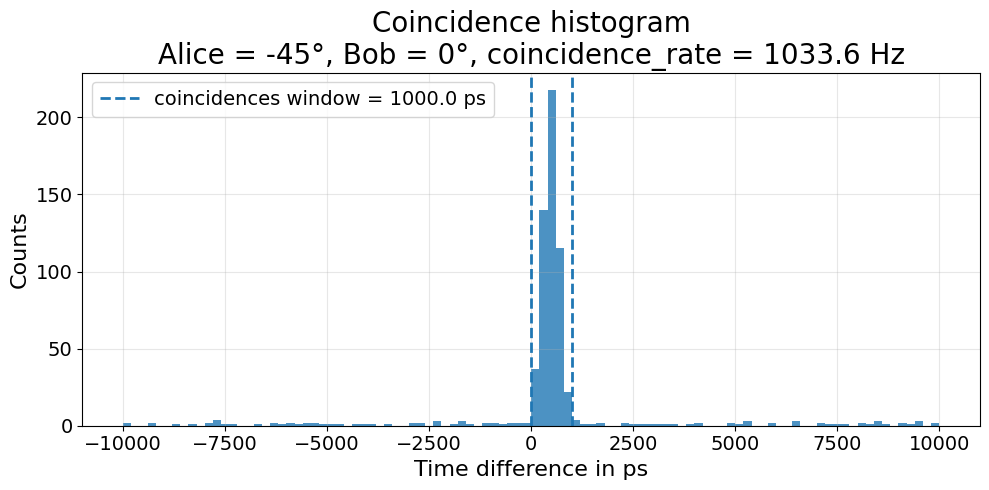}
    \end{minipage}\hfill
    \begin{minipage}[t]{0.48\textwidth}
        \centering
        \includegraphics[width=\textwidth]{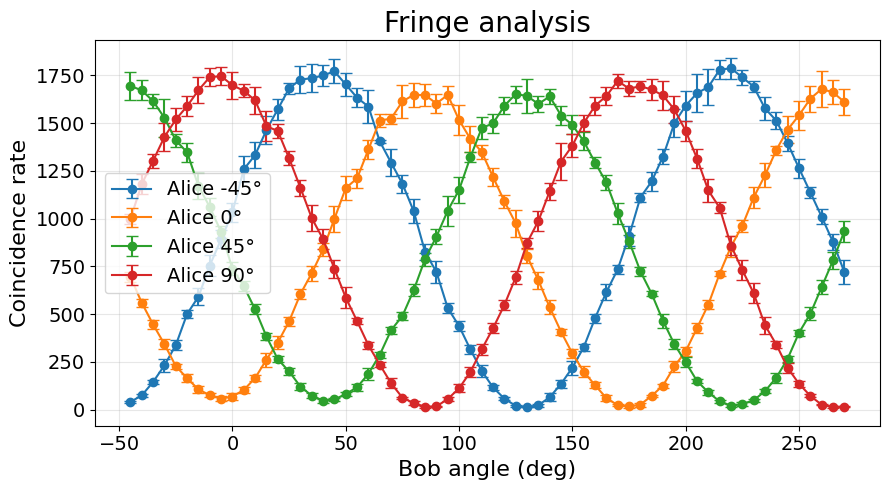}
    \end{minipage}
    \caption{Experiment results. Left: coincidence histogram between Alice and Bob detectors, showing a peak near zero delay after software correction. Right: polarization correlation fringe from the automated sweep, with measured visibility of 98 percent.}
    \label{fig:combined_results}
\end{figure*}

We first ran coincidence detection between Alice and Bob to verify distributed acquisition and timing alignment. The coincidence capability retrieved synchronized timestamps from both time tagger agents and computed a histogram with a 1 nanoseconds coincidences window. In Fig.~\ref{fig:combined_results}, the left panel shows a clear peak near zero delay after software correction. Without this correction, the coincidence peak was initially offset by approximately $7.1\,\mu$s due to path and system delays, and was realigned in software by compensating for this offset.

This type of measurement across remote locations traditionally requires recording data locally at each location and performing offline post-processing to identify coincidences, often concluding \emph{after the fact} that entanglement was present. In contrast, our approach enables online coincidence analysis across distributed nodes, allowing immediate observation and validation of entanglement during experiment execution.

We then ran an automated polarization-fringe workflow. Alice used four fixed analyzer settings ($-45^\circ$, $0^\circ$, $45^\circ$, and $90^\circ$), while Bob was swept over a range of angles. For each pair of rotation settings, the coordinator invoked the \textit{polarization analysis capability}, which used the \textit{polarization controller resource agents} and the \textit{coincidence analysis capability} to produce one measurement point. Fig.~\ref{fig:software_interraction} shows the software interaction flow. In Fig.~\ref{fig:combined_results}, the right panel shows the resulting fringe with about 98 percent visibility.

Together, these results confirm that the Measurement Plane can coordinate remote devices and automate multi-step characterization experiments, while enabling live feedback that is typically not available in conventional post-processed workflows.

\subsection{Experiment Automation}
The deployment also showed practical workflow gains. Instead of coordinating many manual steps and custom scripts for each run, users submit one experiment description and let the coordinator execute it. In our setup, this reduced experiment-specific logic from hundreds of script lines to a short workflow specification, improved consistency across repeated runs, and significantly reduced experiment setup and execution time.

\section{Conclusion and Perspectives}
\label{sec:conclusion}
In this work, we present the Measurement Plane, which separates user access, workflow coordination, measurement logic, and hardware interfaces into independent layers. This structure enables measurement capabilities to be reused across experiments while keeping user workflows decoupled from device-specific implementations.

Our deployment across two nodes showed that the framework can coordinate remote resources for quantum-network characterization. The coincidence and polarization-fringe results demonstrate that the architecture supports automated and reproducible multi-step experiments on a real testbed.

This work demonstrates feasibility rather than scale limits. Future work will include scalability benchmarking, broader capability coverage, and tighter integration between physical and simulated testbeds. In particular, physics-informed quantum-network simulation can model realistic testbed components, including sources, detectors, resources, and measurement procedures \cite{amlou2026physics}. Combining such simulation tools with the Measurement Plane would enable a unified framework in which the same measurement or experiment description could be executed either on virtual resources or on the physical testbed when available. This direction aligns with ongoing efforts that emphasize the importance of digital-twin capabilities for quantum networks and quantum internet infrastructure, enabling experiment preparation, validation, and system-level evaluation before real deployment \cite{doolittle2026quantum, irtf-qirg-qi-multiplane-arch-01}.

\section*{Disclaimer} 
Any mention of commercial products or references to commercial organizations is for information only; it does not imply recommendation or endorsement by NIST, nor does it imply that the products mentioned are necessarily the best available for the purpose.

\bibliographystyle{IEEEtran}
\bibliography{references}

\end{document}